# Quantitative precipitate classification and grain boundary property control in Co/Ni-base superalloys

**T. P. McAuliffe**[1], I. Bantounas[1], L. R. Reynolds[1], A. Foden[1], M.C. Hardy[2], T. B. Britton[1] & D. Dye[1]

[1] Imperial College London, Department of Materials, Prince Consort Road, Imperial College London, London, SW7 2AZ, UK
[2] Rolls-Royce plc, PO Box 31, Derby DE24 8BJ, UK

## Abstract

A correlative approach is employed to simultaneously assess structure and chemistry of (carbide and boride) precipitates in a set of novel Co/Ni-base superalloys. Structure is derived from electron backscatter diffraction (EBSD) with pattern template matching, and chemistry obtained with energy dispersive X-ray spectroscopy (EDS). It is found that the principal carbide in these alloys is Mo and W rich with the $M_6C$ structure. An $M_2B$ boride, also exhibiting Mo and W segregation is observed at B levels above approximately 0.085 at.%. These phases are challenging to distinguish in an SEM with chemical information (EDS or backscatter Z-contrast) alone, without the structural information provided by EBSD. Only correlative chemical and structural fingerprinting is necessary and sufficient to fully define a phase. The identified phases are dissimilar to those predicted using ThermoCalc. We additionally perform an assessment of the grain boundary serratability in these alloys, and observe that significant amplitude is only obtained in the absence of pinning intergranular precipitates.

## 1 Introduction

Engineering grain boundary character is essential to optimisation of superalloy microstructure and performance. Toward this endeavour we require precise knowledge of chemistry, distribution, and structure of intra/intergranular precipitates, particularly carbides and borides. These phases form through all stages of alloy processing, from initial casting through to final ageing treatments. In the literature, refractory-rich grain boundary precipitates may be referred to as a 'carbide' without any effort to differentiate between $M_{23}C_6$, $M_6C$ or boride structures. Due to preferential elemental segregation of high Z-number elements, these all exhibit high backscatter scanning electron microscope (SEM) contrast. There is significant evidence that these precipitates' exact character has a significant effect on mechanical and environmental stability, especially in high temperature applications, despite their small volume fraction [1–4]. A secondary effect of grain boundary precipitate interactions is the ability to facilitate grain boundary serration mechanisms, also an essential consideration in modern superalloy grain boundary engineering [5,6].

This work develops the V208 series first presented by Knop et al [7–9]. A set of Co/Ni-base superalloys based on V208C are presented, with Mo additions (for solid solution strength) and varied C, B, Zr, and Ti content for grain boundary chemistry adjustment. We additionally characterise as-received coarse-grained RR1000 as well as cast and wrought V208C for comparison. Intergranular precipitates are quickly and accurately classified using a correlative electron backscatter diffraction (EBSD) / energy dispersive X-ray spectroscopy (EDS) approach [10]. We examine the robustness of the classification, present chemical statistics as a function of precipitate structure, and compare our observations to thermodynamic modelling. We also briefly investigate the effect of intergranular phase morphology on grain boundary control. We show that in this alloy series boundaries may only be serrated in the absence of spatially dense precipitate boundary coverage, regardless of cooling rate from solution.

## 2 Background

In this study we develop and process a new series of Co/Ni-base superalloys with varied C, B, Zr and Ti compositions to adjust grain boundary chemistry [11]. We then observe and discuss the formation of various intergranular phases and grain boundary morphologies upon independently varied heat treatment. We also characterise and compare as-received coarse-grained RR1000 and cast V208C. To provide background, here we briefly review the metallurgy of phases within superalloys including their influence on grain boundary serration, and modern approaches to accurate and statistically robust phase characterisation.

### 2.1 Carbide & boride precipitation

A wide variety of carbide and boride structures are known to precipitate in superalloys, including but not limited to MC, $M_{23}C_6$, $M_6C$, $M_2C$, $M_2B$, $M_3B_2$ and $M_5B_3$, where M is a dominant metallic enrichment, generally a refractory element. A selection of prior studies within superalloys and steels with similar refractory contents are presented in Table 1. Current thinking is that a significant mismatch between solute atomic radii to the average of the pseudo-FCC ($\gamma$ / $\gamma'$) matrix (smaller for C and B, larger for refractories) leads to a driving force for these elements to migrate to grain boundaries. Here the strain dipoles accommodating lattice parameter mismatch from interstitial or substitutional solutes can be relieved. The migrated elements tend to combine, forming intermetallic and ceramic compounds such as topologically close packed (TCP) phases, carbides and borides.

MC carbides generally precipitate at high temperatures (for example over 1100°C in Udimet-520 [12]), during casting ('primary' MC carbides) or homogenisation treatments ('secondary'). These are stable at high temperature, and are difficult to dissolve once formed without risking incipient melting. MC carbides may exhibit significant variability in composition due to forming over a wide temperature range during solidification [13]. Their relative stability also means they are a common decomposition product of other carbides, such as $M_2C$ [13–16], with the associated refractory rejection possibly also leading to TCP phase precipitation [17]. In superalloys their presence is generally welcomed, with certain systems utilising them during supersolvus forging, providing hot ductility, or for the prevention of hot zone cracking [18].

| Type | System | 'M' Enrichment | Material | Technique | Ref |
|---|---|---|---|---|---|
| MC | Superalloy | Ti | René 88DT | XRD | [44] |
| | | Nb | Inconel 718 | TEM-EDS | [32] |
| | | Ti, Mo | Udimet 520 | EDS | [12] |
| | | Ta | STAL-15CC | APT | [42] |
| | | Ti, Ta, Nb, Mo | ME3 | STEM-EDS | [46] |
| | | Zr, Ta | Co/Ni superalloy | EBSD+EDS | [10] |
| | Steel | Mo, V | AISI M2 HSS variants | APFIM | [47] |
| | | Mo, V, W | AISI M2 HSS variants | TEM-EDS | [14] |
| | | Nb, Ta, V | Ferritic/Martensitic Steel | STEM-EDS | [48] |
| | | V, Cr, Mo | AISI M2 HSS variants | EDS | [15] |
| $M_{23}C_6$ | Superalloy | Cr, Mo | RR1000 | STEM-EDS | [43] |
| | | Cr, Mo | Udimet 720 | TEM-EDS | [49] |
| | | Cr, Mo | Udimet 520 | EDS | [12] |
| | | Cr | Inconel 738 | STEM-EDS | [50] |
| | | Cr, Mo | STAL-15CC | APT | [42] |
| | | Cr, Mo, W | ME3 | TEM-EDS | [46] |
| | Steel | Cr, Co | Ferritic/Martensitic Steel | STEM-EDS | [48] |
| | | Cr | CrMnFeCoNi HEA | STEM-EDS | [51] |
| $M_6C$ | Superalloy | W, Mo | K465 superalloy | STEM-EDS | [52] |
| | Steel | W, Mo | AISI M2 HSS variants | STEM-EDS | [13] |
| | | Mo, W | AISI M2 HSS variants | STEM-EDS | [14] |
| | | Fe, Co | Ferritic/Martensitic Steel | STEM-EDS | [48] |
| $M_2C$ | | Mo, Cr | Hastelloy N | EDS | [19] |
| | | Mo, Cr | AISI M2 HSS variants | APFIM | [47] |
| $M_3C$ | Steel | Co | Ferritic/Martensitic Steel | STEM-EDS | [48] |
| $M_5B_3$ | Superalloy | Cr, Mo, W | René 88DT | XRD | [44] |
| | | Cr, Mo, W | STAL-15CC | APT | [1] |
| | | Cr, Mo | In 738 | STEM-EDS | [53] |
| | | Cr, Mo | STAL-15CC | APT | [42] |
| | | Cr, Mo, W | ME3 | TEM-EDS | [46] |
| $M_3B_2$ | Superalloy | Cr, Mo, W | René 88DT | XRD | [44] |
| | | Nb, Mo, Cr | Inconel 718 | EDS | [45] |
| $M_2B$ | Steel | Cr, Fe | Austenitic stainless steel | STEM-EDS/EELS | [54] |
| | | Cr, Fe | AISI 1045 steel | TEM-EDS | [55] |
| | | Cr, Fe | 18/20 Ni/Cr stainless steel | XRD | [56] |

**Table 1**: Carbides and borides commonly observed in superalloys, with the corresponding refractory affinities measured with a variety of characterisation techniques[1].

$M_{23}C_6$ and $M_6C$ phases often form on grain boundaries at intermediate temperatures (700 – 1000°C), (unfortunately) coinciding with optimum ɣ' ageing regimes. They may also form from the decomposition of other carbides such as $M_2C$ or $MC$, and on occasion precipitate intragranularly [19,20]. As we have presented in Table 1, $M_6C$ tends to exhibit a greater W affinity than $M_{23}C_6$, so unsurprisingly is observed in Co-base superalloys that require high atomic fractions of W [20–22].

## 2.2 Grain boundary serration

Heat treatments designed to mechanically lock grain boundaries together have been used since first being reported for superalloys in 1976 by Miyagawa et al [23] and Larson [24], though the morphology was reported in austenitic stainless steels 10 years prior [25]. They universally involve slow cooling through an intergranular phase solvus temperature. There is significant evidence that engineering such microstructures improves creep ductility and life, where intergranular cracking and cavitation modes are exhibited [6,23–25]. Serrations directly affect the character of strain evolution in high temperature deformation regimes. *In-situ* digital image correlation (DIC) studies on René-104 have shown that serrations reduce strain concentration around microstructural features such as annealing twin (Σ3) boundaries, and plasticity is distributed more evenly across the microstructure [26,27]. This is as originally suggested by Larson [24] and later validated by Carter *et al* [26] who showed that strain concentration fields at serrated grains boundaries are smaller in magnitude and extend further into grain interiors than for non-serrated. This is likely due to a reduction in the accommodation of strain through grain boundary sliding (GBS), which reduces the onset of tertiary creep.

The processes by which grain boundaries serrate during heat treatment are still the subject of some debate. The dominant mechanism varies between alloy systems as a function of ɣ' and intergranular phase solvus temperatures. Larson [24] showed that an air cool through the ɣ' solvus develops a serrated microstructure in Inconel 792, with the mechanism attributed to ɣ' intergranular nucleation followed by subsequent boundary migration. Highly serrated boundaries are observed to only exhibit large, globular carbides. This is attributed to a preference for this morphology to precipitate at serration nodes during intermediate temperature ageing. Lower temperature ageing prior to air cooling produced smoother grain boundaries with film-like $M_{23}C_6$ carbides. The inference drawn at the time was that serrated boundaries promoted globular $M_{23}C_6$. An alternative, non-mutually exclusive interpretation is that the film-like $M_{23}C_6$ carbides, if aged in to the microstructure at intermediate temperature (even if above the ɣ' solvus), prevent serration. Miyagawa *et al* [23] explain their similar observations (after comparable heat treatments to Larson) through nucleation and resulting preferential growth of $M_{23}C_6$ precipitates during cooling. In this alternative scheme, carbide growth into the grain bulk along a preferred crystallographic direction provides a lower interfacial energy plane for the grain boundary to migrate to.

The ɣ'-driven mechanism proposed by Koul & Gessinger [5] involves preferential intergranular precipitation due to superlattice misfit relief. They argue that the strain energy difference between the boundary and matrix facing sides of the ɣ' particle provides a driving force for its migration in the direction of the boundary normal, until this energy is matched by the boundary line tension. For this mechanism to operate effectively the grain boundary carbide ($M_{23}C_6$ in their study) solvus temperature should to be lower than that of ɣ'. The key requirement is that grain boundary segments between ɣ' particles must be mobile during the extended nucleation period. Consequently ɣ' and carbide-driven mechanisms cannot operate simultaneously: if a precipitate is already pinning the boundaries, slow precipitation of a lower temperature phase will not successfully serrate the boundaries (regardless of whether this is ɣ' or carbide). This effect is presented in Nimonic 105, where $M_{23}C_6$ precipitation precedes ɣ' and does not itself serrate the boundaries. The pinning precipitates then prevent ɣ' from subsequently doing so.

There are now many observations of carbide induced serrations, including the 'zig-zag' morphology reported by Miyagawa et al [23] and Yamazaki [25] in stainless steels early on, as well as more recent observations in non-ɣ' containing Ni-base alloys [6]. However, there has been substantially less effort to verify any carbide-based serration mechanisms further than reported by Yamazaki in non-ɣ' containing Ni alloys and steels. This remains an interesting question, yet to be validated with modern microstructural characterisation techniques, but is beyond the scope of the present study due to the presence of large volume fractions (50-55% in V208C) of ɣ'.

## 2.3 Phase classification

Accurate phase identification requires knowledge of precipitate crystal structure as well as chemistry. Here we describe phase classification in terms of assigning a crystal with specific structure and chemistry to a common label or class. In many cases, classification can be performed with only chemistry or only structure if the domain of the problem is constrained (e.g. there is knowledge of the thermodynamics and kinetics for microstructural formation). This is different to phase identification, where the structure and chemistry of the phase is unknown. To perform accurate classification, we must sample both the chemistry and structure.

Atom probe tomography (APT) provides unparalleled chemical resolution, but is site specific and spatially imprecise in the radial directions of the ionised tip [28]. Typically this technique samples chemistry with nm spatial resolution, but recent methods have improved local structural measurements based upon matching to a library of potential detector hit maps [29].

Conventional transmission electron microscopy (TEM) is the traditional method for structure determination, and relies on analysis and indexing of selected area diffraction patterns. Spatial resolution of this approach is limited by the size of the selected area aperture employed, unless overlapping diffraction patterns can be successfully deconvolved. In practice, after (usually challenging) sample preparation a nearby zone axis is located and a two-dimensional projection of the sampled region's reciprocal lattice is measured. These can be compared to (evenly sampled in orientation space) libraries of kinematically or dynamically simulated spot patterns for candidate crystal structures, but there may be cases of pseudosymmetry and strong pattern similarity, especially between phases with near identical structure. This is confounded by the fact that upon rotation to a zone axis one only ever samples two coplanar reciprocal lattice dimensions. This leads to a 180°

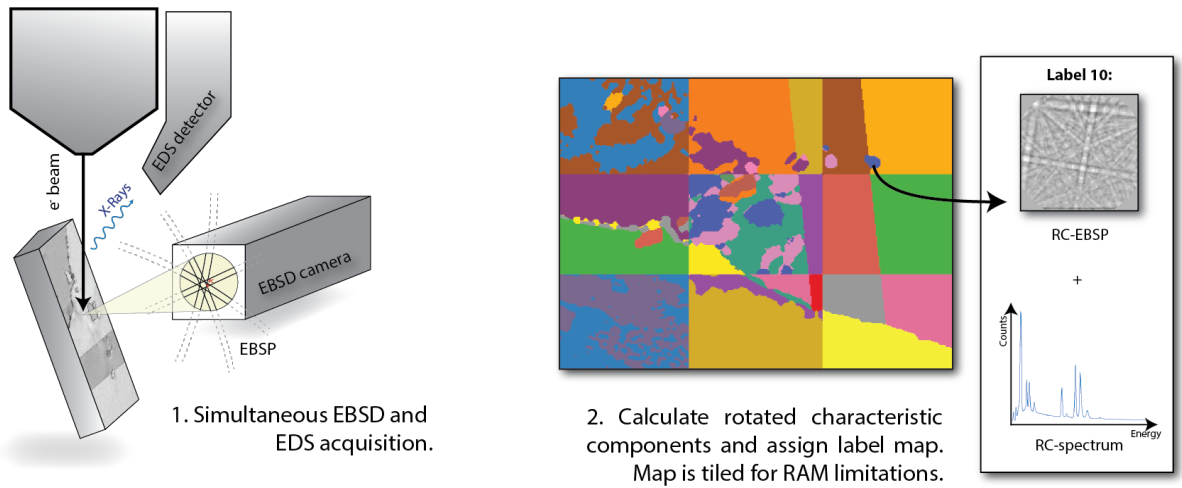

1. Simultaneous EBSD and EDS acquisition.

2. Calculate rotated characteristic components and assign label map. Map is tiled for RAM limitations.

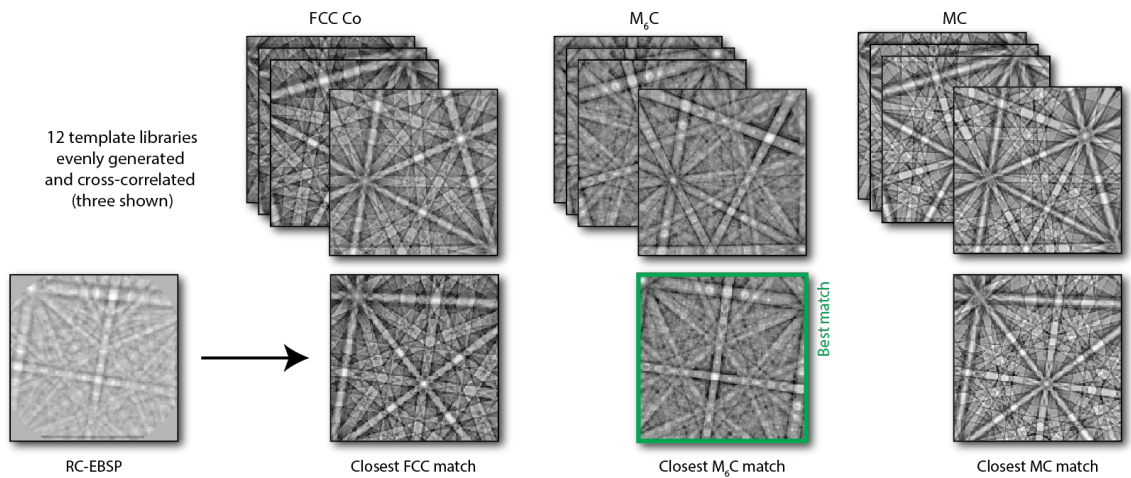

3. RC-EBSP cross-correlated in Fourier space with dynamically simulated templates for structure ID.

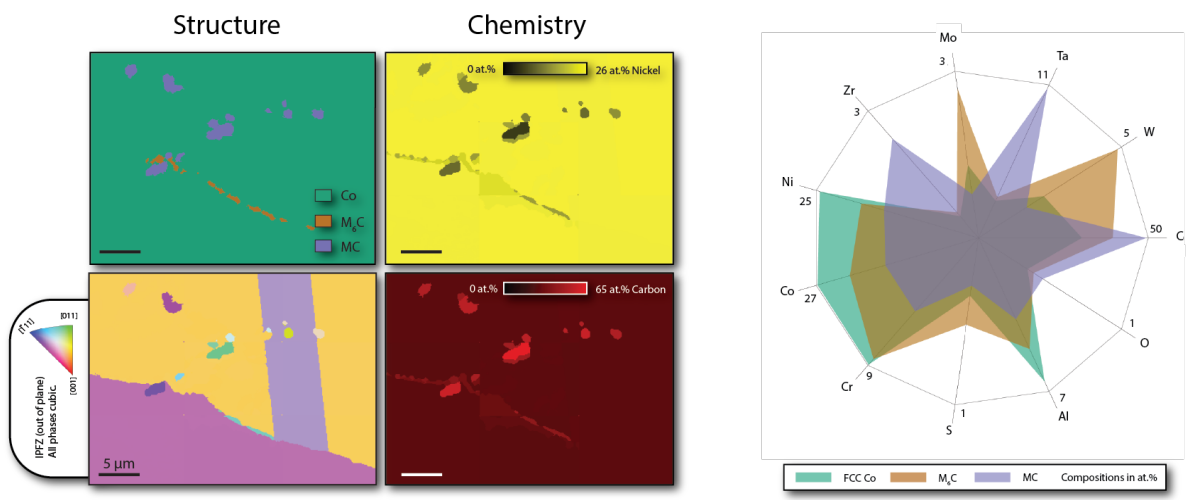

4. RC-spectra independently quantified, and structure/chemistry relationships and inferences drawn.

**Figure 1**: The EBSD/EDS PCA approach. Chemical and structural information is combined and corresponding relationships are extracted using statistical methods from the dataset.

| Alloy | Co | Ni | Mo | Cr | Al | W | Ta | C | B | Zr | Ti |
|---|---|---|---|---|---|---|---|---|---|---|---|
| 1 | ~36 | ~36 | 2 | 12 | 10 | 2.75 | 1.25 | **0.300** | **0.085** | 0.040 | 0 |
| 2 | ~36 | ~36 | 2 | 12 | 10 | 2.75 | 1.25 | **0.296** | **0.043** | 0.040 | 0 |
| 3 | ~36 | ~36 | 2 | 12 | 10 | 2.75 | 1.25 | **0.224** | **0.041** | 0.040 | 0 |
| 4 | ~36 | ~36 | 2 | 12 | 10 | 2.75 | 1.25 | **0.100** | **0.042** | 0.040 | 0 |
| 5 | ~36 | ~36 | 2 | 12 | 10 | 2.75 | 1.25 | **0.100** | **0.020** | 0.040 | 0 |
| 6 | ~36 | ~36 | 2 | 12 | 10 | 2.75 | 1.25 | **0.180** | **0.085** | 0.040 | 0 |
| 7 | ~36 | ~36 | 2 | 12 | 10 | 2.75 | 1.25 | **0.250** | **0.125** | 0.040 | 0 |
| 8 | ~36 | ~36 | 2 | 12 | 10 | 2.75 | 1.25 | **0.250** | **0.200** | 0.040 | 0 |
| 9 | ~36 | ~36 | 2 | 12 | 10 | 2.75 | 1.25 | **0.300** | **0.110** | 0.040 | 0 |
| 10 | ~36 | ~36 | 2 | 12 | 10 | 2.75 | 1.25 | 0.100 | 0.042 | **0.020** | **0.1** |
| 11 | ~36 | ~36 | 2 | 12 | 10 | 2.75 | 1.25 | 0.100 | 0.042 | **0.020** | **0.2** |
| 12 | ~36 | ~36 | 2 | 12 | 10 | 2.75 | 1.25 | 0.100 | 0.042 | **0.020** | **0.3** |
| 13 | ~36 | ~36 | 2 | 12 | 10 | 2.75 | 1.25 | 0.100 | 0.042 | **0.040** | **0.3** |
| V208C [9] | ~36 | ~34 | 0 | **15** | **10.5** | 3 | 1 | 0.150 | 0.200 | 0.040 | 0 |
| L1 | ~34 | ~35 | 2 | **13** | **12** | 2.75 | 1.25 | 0.300 | 0.110 | 0.040 | 0 |
| L2 | ~34 | ~35 | 2 | **13** | **12** | 2.75 | 1.25 | 0.300 | 0.440 | 0.040 | 0 |

**Table 2**: Nominal compositions (in at.%) for the alloys developed in this study, as well as V208C [9]. The first nine alter C and B content. Alloys 10-13 adjust Zr and Ti composition.

ambiguity in crystal orientation: three non-coplanar vectors are required for a full description [30]. However, spot diffraction patterns are extremely difficult to interpret and index if the electron beam is not aligned to a high symmetry zone, unless the full 3D reciprocal lattice is measured with a tomographic approach [31]. Chemical information can be obtained either using energy dispersive X-ray spectroscopy (EDS) for heavier elements [12,15,32] and local bonding can be explored using energy electron loss spectroscopy (EELS) [31,33].

For both APT and TEM, sample preparation is complex, and the volumes explored are small. This limits their use for exploring a wide range of carbides and borides located across a range of different microstructural regions. This motivates us to use a large area mapping method based upon combined electron backscatter diffraction (EBSD) and energy dispersive X-ray spectroscopy (EDS) in the scanning electron microscope (SEM).

Here a correlative EBSD and EDS method is employed, together with statistical methods, to enable analysis of large area data sets. EBSD involves the measurement of electron backscatter diffraction patterns (EBSPs). These are formed from near elastic diffraction scattered electrons, and correspond to projections of the lattice planes [34,35]. EBSPs are formed from a relatively large solid angle of diffracted electrons, so more than three geometrical conditions (in this case Kikuchi bands corresponding to lattice planes) are normally sampled, as required to determine a unique orientation solution. Conventionally a Hough (Radon) transform and set of interplanar angle lookup tables are employed to index crystallographic planes to determine phase and orientation.

Here we adopt an informatics-type approach, combining EBSD and EDS information into large data matrices from which we extract the strongest (correlated) structural and chemical signals using unsupervised machine learning (principal component analysis, PCA). These signals are the principal components of the data matrix. A VARIMAX rotation is performed on the principal components and corresponding (spatial) scores, as per Wilkinson *et al* [36] and with including the EDS signal as per McAuliffe *et al* [10]. This rotation acts to maximise and equalise the variance of the principal components. We then enforce positivity on the rotated components, mimicking the statistics of experimentally measured EBSPs to obtain physically meaningful results. The procedure generates rotated characteristic components (RCCs), in turn containing rotated characteristic EBSPs (RC-EBSPs) and spectra (RC-spectra). These can be independently indexed and quantified with higher confidence as they haveenhanced signal to noise as compared to the raw data. The approach lets us reduce a (for example) 40,000 scan point map, each with 200-by-200 EBSP pixels and 2048 EDS energy bins down to as few RCCs as there are grains (or sub-grains if we permit oversampling) in an area of interest (AOI). At intermediate magnification this value is usually 50 – 100 in our relatively coarse grained materials. This approach is presented and validated in further detail in our prior work and is schematically presented in Figure 1.

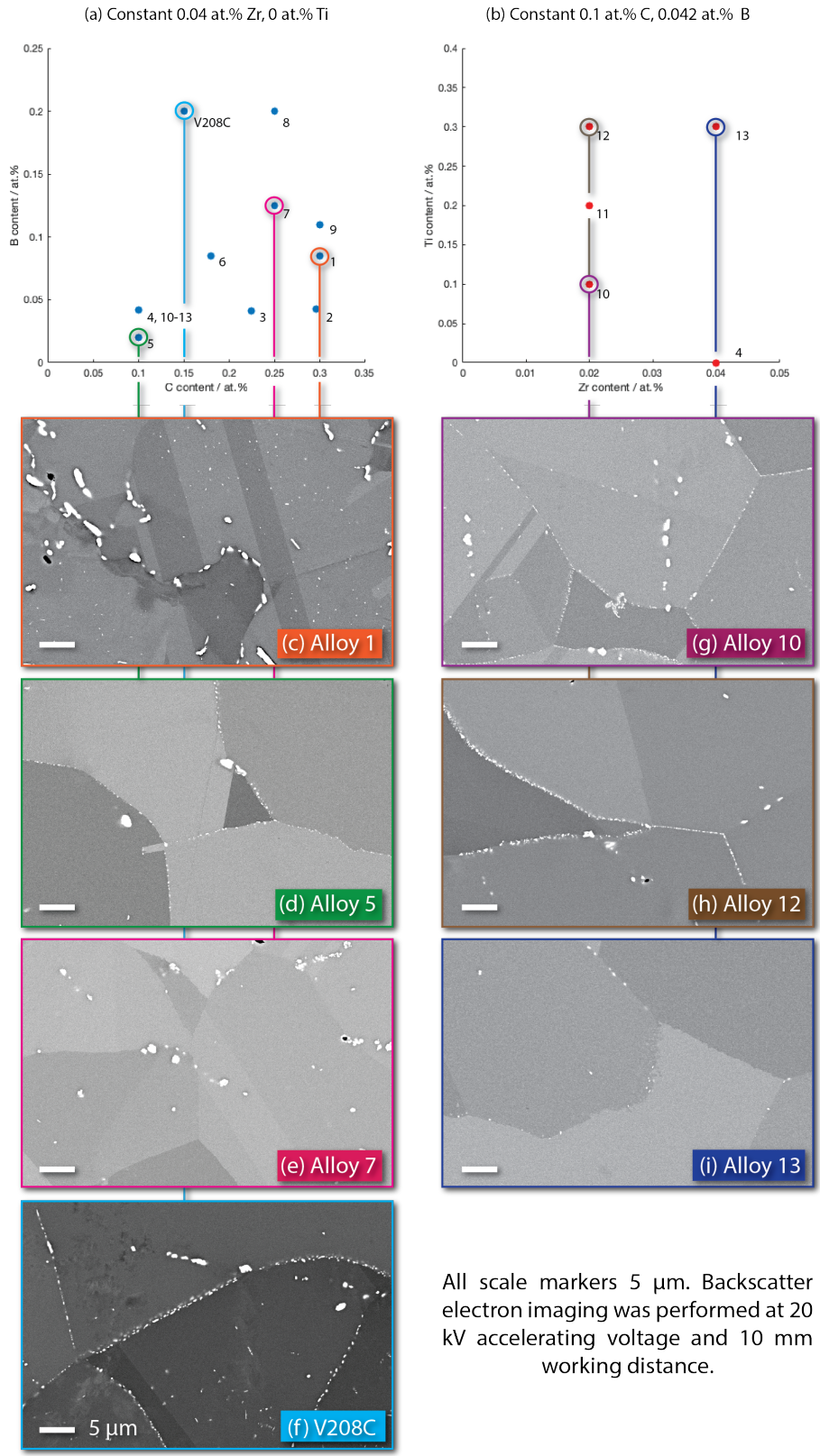

**Figure 2**: Nominal compositions (in at.%) for the alloys developed in this study. Alloys 1-9 and V208C (a) have constant 0.040 at.% Zr and 0 at.% Ti. Alloys 4, 10-13 (b) have constant 0.1 at.% C and 0.042 at.% B. (c)-(h) present representative BSE images of grain boundary morphology for a highlighted sub-selection of the alloys.

Combining structure assignment and chemical measurement lets us classify the phase of a labelled region (usually corresponding to a single precipitate grain). We can then gather statistics of chemistry and structure as a function of the other, for example. In this work, we classify structure via the refined template indexing (RTI) procedure developed by Foden *et al* [37] using dynamically simulated libraries of template EBSPs for each candidate phase, also depicted in Figure 1. The candidate phases may be selected from observations in previous studies, or via computational predictions (e.g. with ThermoCalc). The simulated library EBSPs are cross-correlated in Fourier space with the RC-EBSPs. A scan point is assigned to the structure and candidate orientation with the highest cross-correlation peak height. Subsequent iterative refinement determines the precise misorientation of the measured pattern to that of its best matching template. This approach allows us to distinguish phases of similar structure that share many Kikuchi band features, especially since many Radon transform based indexing approaches only consider up to a dozen interplanar angles, which for similarly symmetric structures may be shared. Distinguishing pseudo-FCC (ɣ / ɣ') matrix from MC carbide with conventional EBSD, for example, is challenging.

*3 Experimental methods*

Sixteen approximately 415 g ingots were made by vacuum arc melting. Compositions are presented in Table 2 and Figure 2. Each was vacuum homogenised for 48 h at 1250°C. Ingots were then hot rolled at nominally 1250°C, with 10-15% reductions from 23 mm square cross-section down to 14 mm. Samples for heat treatment and microstructural characterisation were extracted from the rolled bar. Alloy samples were encapsulated in quartz tube backfilled with Ar for heat treatment; times and temperatures are presented in Figure 3. The solution stage aims to dissolve all the ɣ' precipitated during hot rolling and uncontrolled cooling, in order to 'reset' the alloy and generate a controlled ɣ' distribution upon ageing. Three different cooling rates were trialled for a subset of the alloys in order to investigate the effect of intergranular precipitation on boundary serratability, discussed in section 4.4. As-received coarse grained RR1000 was also characterised. A standard metallographic polishing procedure was used to prepare samples for microscopy:

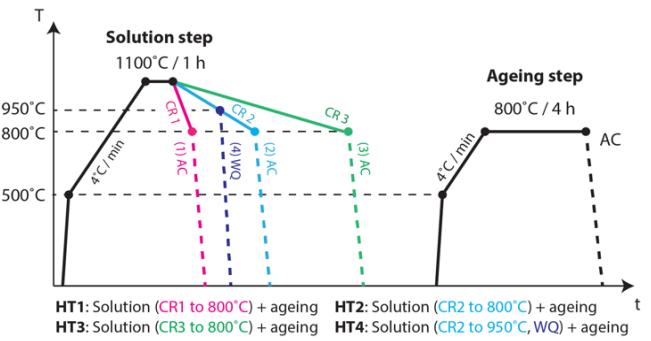

**Figure 3**: Post rolling heat treatments trialled in this study. The solution condition was 1100°C for 1 h, followed by one of: CR1 (20°C/min), CR2 (5°C/min) or CR3 (1°C/min). From 700°C alloys were air cooled, or water quenched at 950°C. All samples were aged at 800°C for ɣ' nucleation and growth.

400 through to 4000 SiC grit grinding, followed by a 1 h neutralised colloidal silica polish.

SEM, EBSD and EDS were performed on a Zeiss Gemini Sigma300 FEGSEM, equipped with Bruker e⁻Flash$^{HD}$ and XFlash 6160 EBSD and EDS detectors respectively. Bruker DynamicS was used to dynamically simulate library EBSPs for each candidate structure [38,39]. From within the fundamental zone for each phase, a SO(3) sampling frequency of 7° was employed for generation of an EBSP template library in the detector reference frame (pattern centre selected with Bruker Esprit 2.1 from the well-indexed matrix regions and simulated patterns sampled as 200 by 200 px). RC-spectra were quantified with Bruker Esprit 2.1 using a P/B ZAF correction algorithm accommodating the 70° sample tilt required for EBSD.

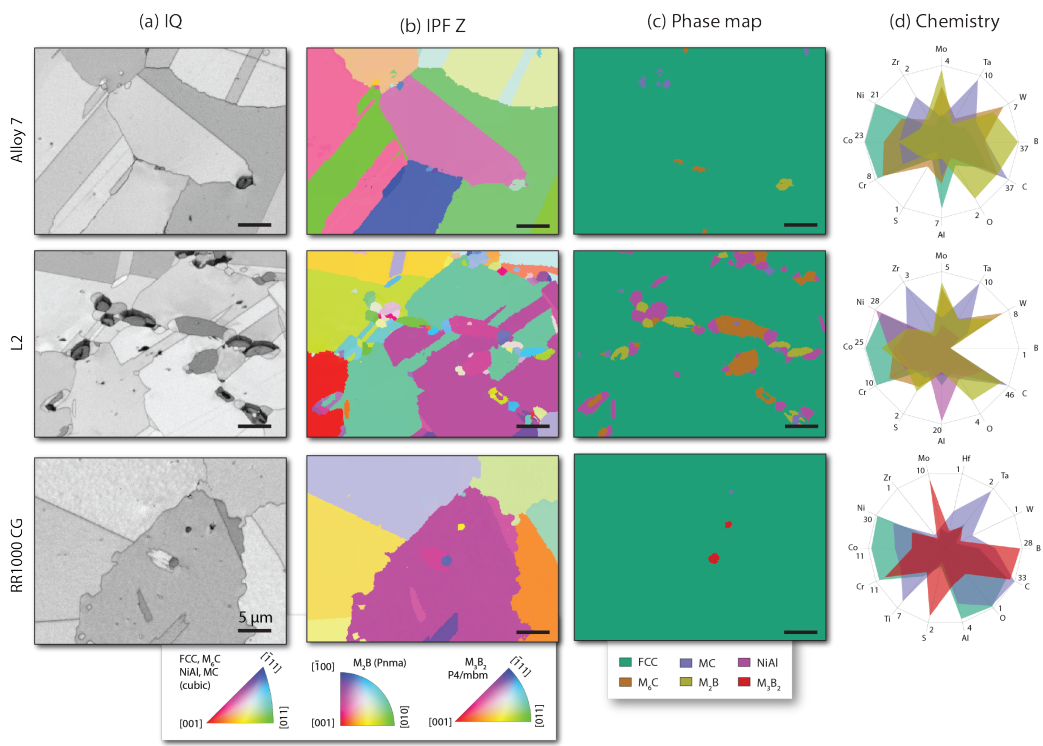

**Figure 4**: Pseudo-backscatter image (a), IPF-Z, Z out of plane (b), phase map (c) and chemical distribution (d) for three selected alloys.

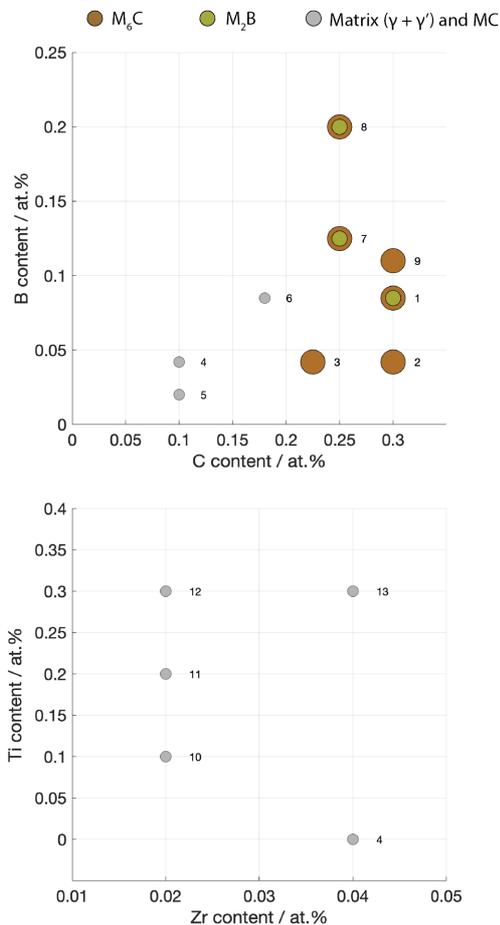

**Figure 5**: Phase diagram of precipitate observations across investigated composition space for alloys 1- 13.

ThermoCalc was used to model the alloys' predicted phase composition. The TCNi-8 database was used.

## 4 Results

### 4.1 Grain boundary precipitation

Combinatorial PCA-EBSD/EDS was used to classify precipitates that nucleated on grain boundaries. EBSP template libraries for candidate structures were generated with Bruker DynamicS: FCC Co/Ni matrix; NiAl and $Co_3W$ intermetallics; $M_6C$, $M_{23}C_6$ and MC carbides; MB, $M_3B_2$ and $M_5B_3$ borides; eta ($Ni_3Ti$), sigma, mu and P topologically close packed phases. The PCA approach then reduces large scans (often with over 40,000 points) down to representative patterns and spectra, with one corresponding to each grain (or sub-grain depending on the extent of oversampling). These are cross-correlated with template libraries for each candidate structure to identify crystallographic phase. The template with the largest correlogram peak height out of all structures' libraries is identified as the best match. Example RCC, phase assignment, and phase specific chemical distribution radar plots are presented for alloy 7, L2 and RR1000 in Figure 4. Alloy 7 exhibits precipitation of the MC and $M_6C$ carbides and the $M_2B$ boride. The MC carbide is mainly enriched in Ta and Zr, while the $M_6C$ carbide is W and Mo rich. The $M_2B$ boride is also Mo and W rich, making it difficult to distinguish from the carbide using conventional EDS chemical mapping. This intergranular morphology is typical of alloys 1 -13,. Alloy L2 additionally sees precipitation of the B2 NiAl intermetallic structure on the grain boundaries, which we unsurprisingly identify as being enriched in Ni and Al, and relatively depleted in other elements. We also present an RR1000 dataset, and see precipitation of the MC carbide and $M_3B_2$ boride, as thermodynamically simulated by Hardy *et al* [40]. The MC carbide sees Ta and Ti segregation, while the boride draws Ta and Mo.

We conduct the same analysis for all of the alloys investigated. The main alloy set (Alloys 1-13) shows varied precipitation of $M_6C$ and $M_2B$ across C, B, Zr and Ti composition space, with no observations of NiAl or any of the other eleven candidate phases. The majority of alloys exhibited MC carbides. Observations of $M_6C$ carbide and $M_2B$ boride precipitation are presented in Figure 5.

Each RCC we extract from the dataset contains an RC-EBSP and RC-spectrum. Having classified each scan point to one of our candidate structures (using the EBSP), we can either quantify the average EDS spectrum for each scan point assigned to a given phase label, or independently quantify the RC-spectrum itself. We have previously shown that these give very similar results, as the VARIMAX rotation applied to make the principal component EBSPs physical has a similar effect on the EDS spectra [10]. Presently we have elected to quantify the average spectrum for each label and quantified compositions using Bruker Esprit 2.1. Subsequently we numerically average the at% chemistries of each phase by alloy, and present in Figure 6 trends in phase chemistry across the alloy series.

All observations of $M_2B$ boride and $M_6C$ carbide see enrichment in Mo and W. The MC carbide and 'matrix' are generally depleted in these elements. The boride is depleted in Cr and Al, while the $M_6C$ carbide appears to have greater tolerance for these elements.

### 4.2 Confidence in phase assignment

The utility of the statistical chemistry-by-phase approach is dependent on accurate phase classification. The Fourier space cross-correlation peak height ('RTI peak height') provides a metric for this assignment quality. Higher values imply stronger similarity between test (RC-EBSP) and reference (library) patterns. The template matching process assigns phase and (unrefined) orientation based on the highest RTI peak height across the template libraries. Figure 7 shows average peak height by assigned phase across the alloy series. We observe that in almost every case the assigned phase is significantly higher than the second closest template structure, and outside of standard errors (error in the mean value of RTI peak height for all labels in an area of interest). Whenever a matrix label is assigned, the runner-up phase is consistently the MC carbide. When an $M_6C$ label is assigned, the runner-up template is always the $M_{23}C_6$ structure. To further characterise the accuracy of the phase assignment and demonstrate the viability of the RTI procedure, presented in Figure 8 are examples of a matrix and $M_6C$ assigned RC-EBSPs. Best, second-best and poorly matching template patterns, with corresponding RTI peak heights, demonstrate the separability of the assigned phase from the alternatives.

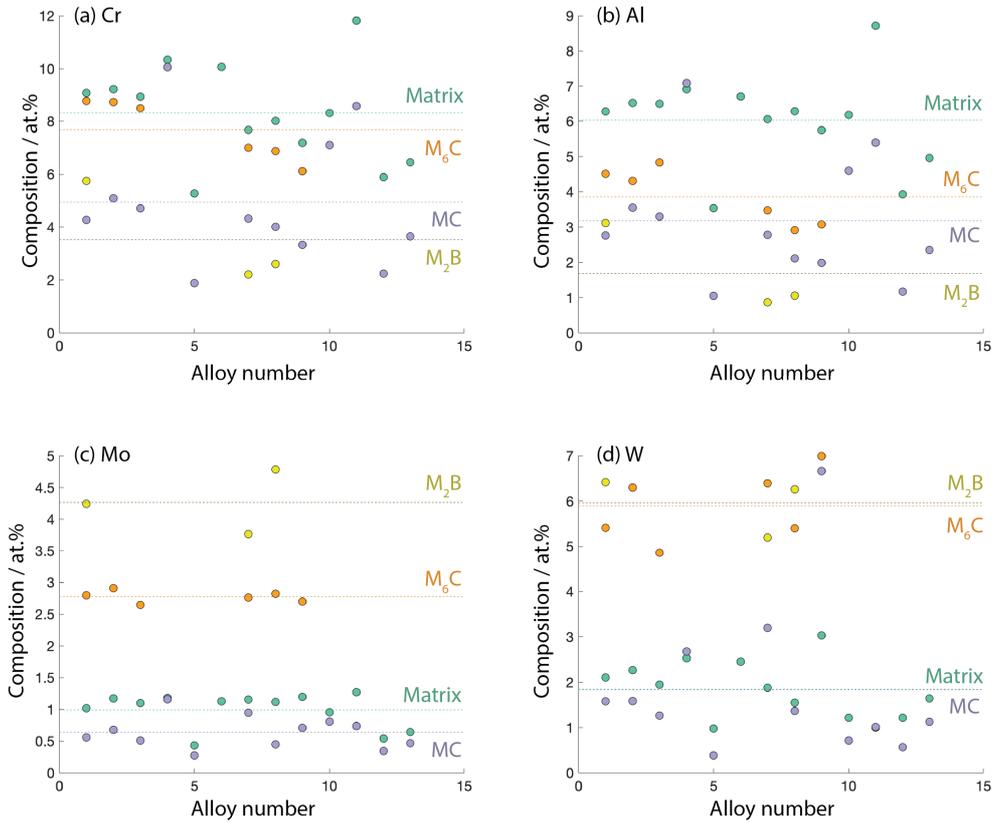

**Figure 6**: Average composition of the M$_2$B boride, M$_6$C carbide, ɣ + ɣ' 'matrix' and the MC carbide across the main alloy series. Cr (c), Al (b), Mo (c) and W (d) are plotted here. Phase composition profiles for the remaining elements are included in the supplementary information.

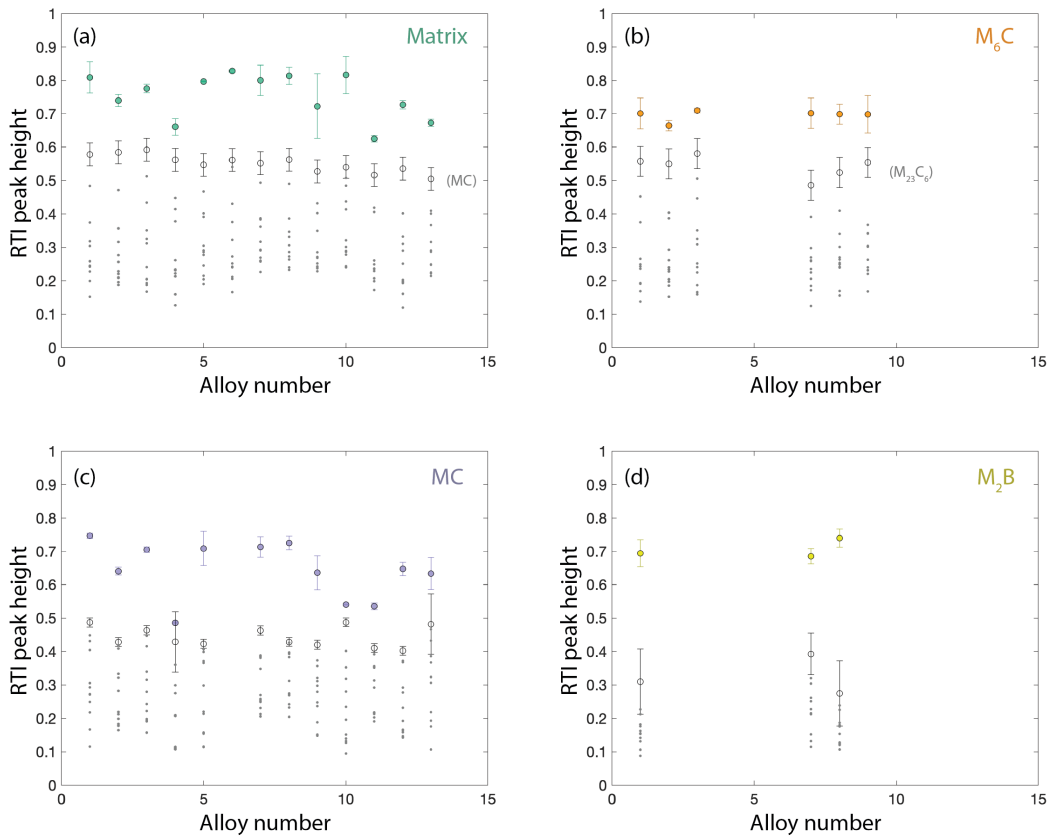

**Figure 7**: Average RTI peak heights (cross-correlation quality) for labels assigned to each of the identified phases, across the main alloy series, with standard-error errorbars. (a) presents the pseudo-FCC 'matrix', (b) the M$_6$C carbide, (c) the MC carbide, and (d) the M$_2$B boride. In each case the second best matching phase is also plotted with standard-error errorbars. This is consistently an MC template for matrix labels, and an M$_{23}$C$_6$ template for M$_6$C labels. There is no consistent runner-up for the MC carbide or M$_2$B boride.

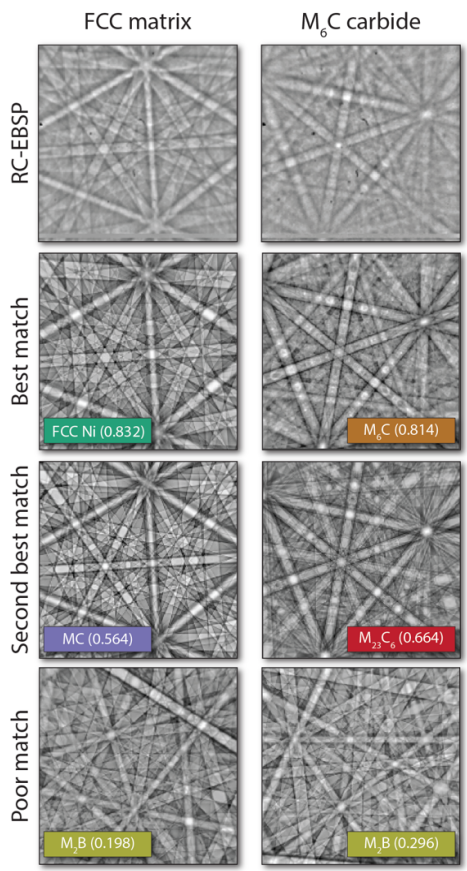

**Figure 8**: RC-EBSPs and the best, second-best and poorly matched simulated template EBSPs. For the FCC 'matrix', the best match (FCC) has an RTI peak height of 0.832. The second closest match, the MC carbide, has a peak height of 0.564 and the 'correct' assignment has a 32% advantage.

MC carbide templates are always runner-up to matrix assignments due to strong similarity in crystal structure. Conventional Hough (Radon) transform EBSP indexing does not account for intra-pattern intensity variation, or presence of minor Kikuchi bands, especially in strongly diffracting crystals which often satisfy the common maximum number to consider for the interplanar angle lookup (often 12). The Fourier cross-correlation handles this well [37], and is able to distinguish the FCC label pattern in Figure 8 from the very similar (in major band trace) MC carbide, with an approximately 32% difference in peak height. Distinguishing the $M_{23}C_6$ and $M_6C$ carbides presents a similar case. There is similarity in the crystal structures leading to systematic relative proximity in peak height. Conventional indexing of this area of interest does not robustly distinguish $M_{23}C_6$ from $M_6C$ structure, and completely misses MC carbides. This issue together with the previously discussed similarity in EDS measured chemistry between $M_6C$ carbide and $M_2B$ boride demonstrates the advantages of a combined approach with independent and accurate quantification / indexing.

### 4.3 Comparison to CALPHAD thermodynamic modelling

Using ThermoCalc with the TC-NI8 database, initially with no structures suspended, predicted phase fractions were calculated for each alloy in the main set. $M_2B$ borides, $M_{23}C_6$ carbides, $Ni_3Ti$ and TCP Mu were predicted (along with an FCC matrix and various populations of L12 ɣ', not plotted). Phase proportions as a function of temperature for alloy 7, whose characterisation is highlighted in Figure 4, are presented in Figure 9.

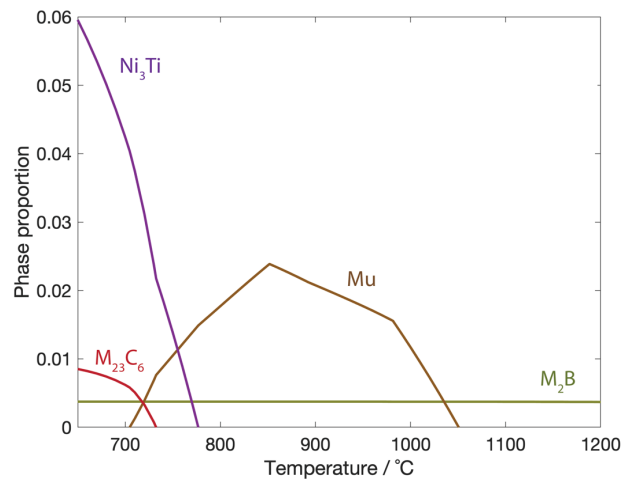

**Figure 9**: Phase proportions of thermodynamically predicted phases (matrix and ɣ' distributions not plotted).

There is a discrepancy between phases predicted by ThermoCalc: $M_2B$, $M_{23}C_6$, mu and eta ($Ni_3Ti$), and those we observe (MC, $M_6C$, $M_2B$). The unobserved phases $M_{23}C_6$, mu and eta were included in the template matching procedure; their template libraries were dynamically simulated and cross-correlated with the label RC-EBSPs but were never matched. As discussed in section 4.2 the RTI phase identification procedure is robust, so we have confidence precipitates have not been mis-identified (for example the $M_{23}C_6$ RTI peak height is consistently lower than that of $M_6C$ for carbide assignment, and there are observable Kikuchi band discrepancies, see Figures 7 and 8).

In Figure 10 (a) and (b) ThermoCalc predicted solvus temperatures are presented for the $M_2B$ boride, this lets us infer how these precipitates are stabilised by variations in C, B, Zr and Ti across the alloy set. Also included are the solvuses for the unobserved $M_{23}C_6$ carbide, (c) and (d). From Figures 10 (a) and (b) we observe that the boride is destabilised (the solvus temperature is lowered) by additions of C, Zr and Ti. Additions of B raise its solvus temperature. A different trend is observed for $M_{23}C_6$. Additions of Ti reduce the carbide's solvus temperature, while Zr appears to raise it (it is not predicted to precipitate at all for alloys with low Zr content). Additions of B do not have a pronounced effect on the predicted stability of $M_{23}C_6$. Comparison to our phase diagram of precipitate observations, Figure 5, partially agrees with the ThermoCalc predictions. In the Zr and Ti enriched alloys we do not observe the boride; it is only observed after about 0.085 at.% B addition.

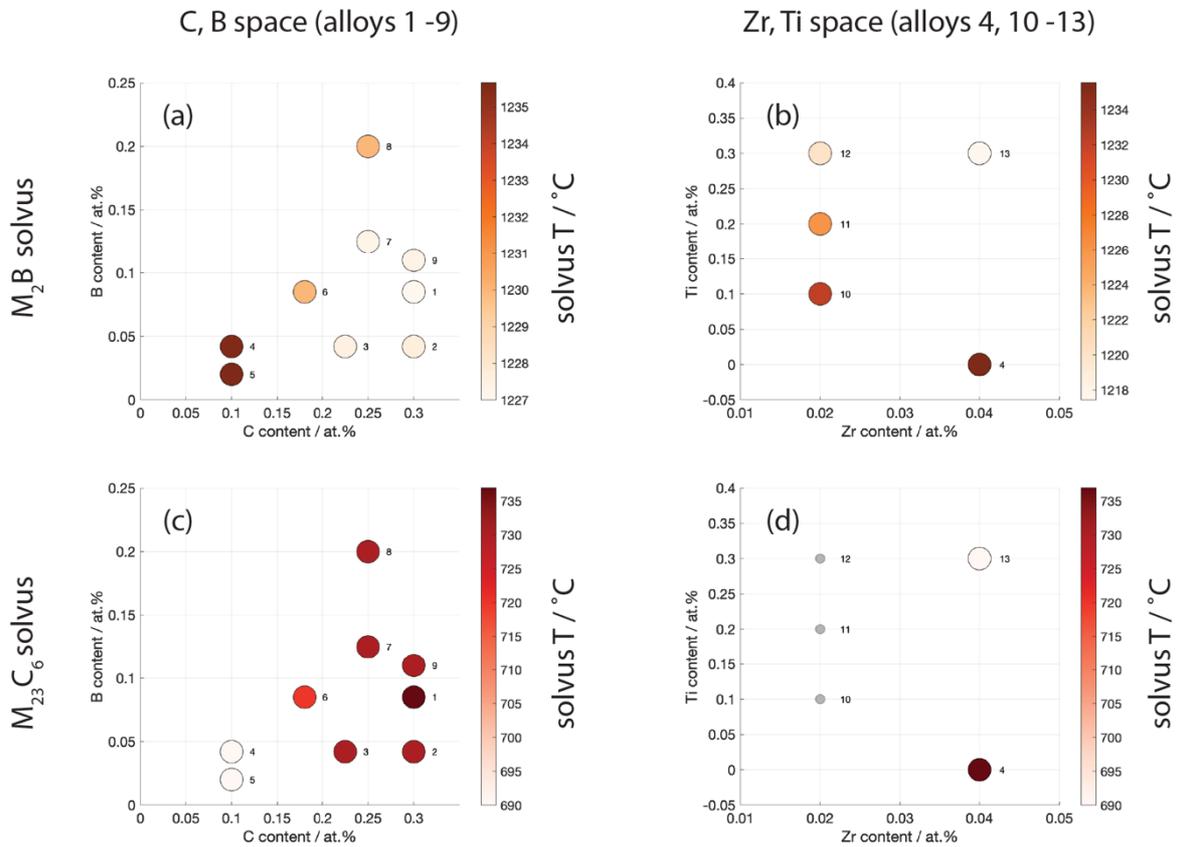

**Figure 10**: ThermoCalc modelled solvus temperatures for the (observed) $M_2B$ boride and (not observed) $M_{23}C_6$ carbide.

### 4.4 Grain boundary serration

Heat treating to produce serrated grain boundaries is known to improve superalloy mechanical performance, especially in deformation regimes where boundary sliding mechanisms are believed to be the principal contributor to strain accumulation [24,26,27,41].

In addition to the standard HT1 employed before characterisation, three further heat treatments were applied to a selection of the alloys in order to promote grain boundary serration. HT2 involved a slow cool through the ɣ' solvus of 5°C / min followed by an air cool from 800°C. HT3 employed an even slower cooling rate of 1°C / min, followed by the same air cool from 800°C. HT4 saw the same 5°C / min cool from solution, but was followed by a water quench at 950°C. All alloys were subsequently aged at 800°C for 4 h to produce an optimal ɣ' distribution. Examination of the heat treated microstructures showed that serrations were possible in this alloy series, albeit of a smaller amplitude than what is often observed, but are completely inhibited by grain boundary coverage of even the finest precipitates. We suggest that in the scheme of Koul & Gessinger [5], the presence of fine boundary particles pins the boundaries, such that upon ɣ' migration there is sufficient line tension to pull back the boundary and prevent extensive serration. In alloy 7, Figures 11(e) and (i), precipitates are sparse (though with greater volume), and serration is readily achievable on many boundaries.

## 5 *Discussion*

Across an even sampling of C, B, Zr, and (low) Ti space, we have processed thirteen alloys (as well as the additional L1, L2, V208C, and RR1000) and characterised the distribution and chemistry of phases within their microstructures with a combinatorial EBSD/EDS approach. In this alloy series we have observed precipitation of intergranular $M_6C$ and MC carbides, and $M_2B$ borides. In alloys with a lower Cr:Al ratio we additionally observe extensive precipitation of the NiAl intermetallic. The observed structures are relatively exotic, with the majority of commercial superalloys exhibiting $M_{23}C_6$ carbides as well as $M_5B_3$ and $M_3B_2$ borides. These structures were all included in the phase-ID stage of our method: evenly SO(3) sampled EBSPs were dynamically simulated for them and compared to our extracted representative patterns. We suggest that the greater Mo and W content of our alloys favours the $M_6C$ carbide and $M_2B$ boride as opposed to $M_{23}C_6$, $M_5B_3$, $M_3B_2$, etc. These structures consistently exhibit strong measured enrichment in these two elements. The apparent consensus from the literature is that $M_{23}C_6$ generally has a greater affinity for Cr than $M_6C$ [42,43]. We have not measured significant enrichment of Cr in $M_6C$ relative to the matrix. Similarly, the $M_5B_3$ and $M_3B_2$ borides have been observed to exhibit greater affinity for Cr than what we see for $M_2B$ [1,44,45]. This likely has significant implications on oxide scale formation and possible ɣ' depletion during high temperature (atmosphere exposed) deformation. Mo and W are not known to be beneficial to stable oxide formation, so precipitates rich in these elements may not be optimal for significant grain boundary coverage. Additionally, Mo and W are known to be slow diffusing elements, which may be an additional factor in our precipitates' ability to inhibit grain boundary serration that we observe.

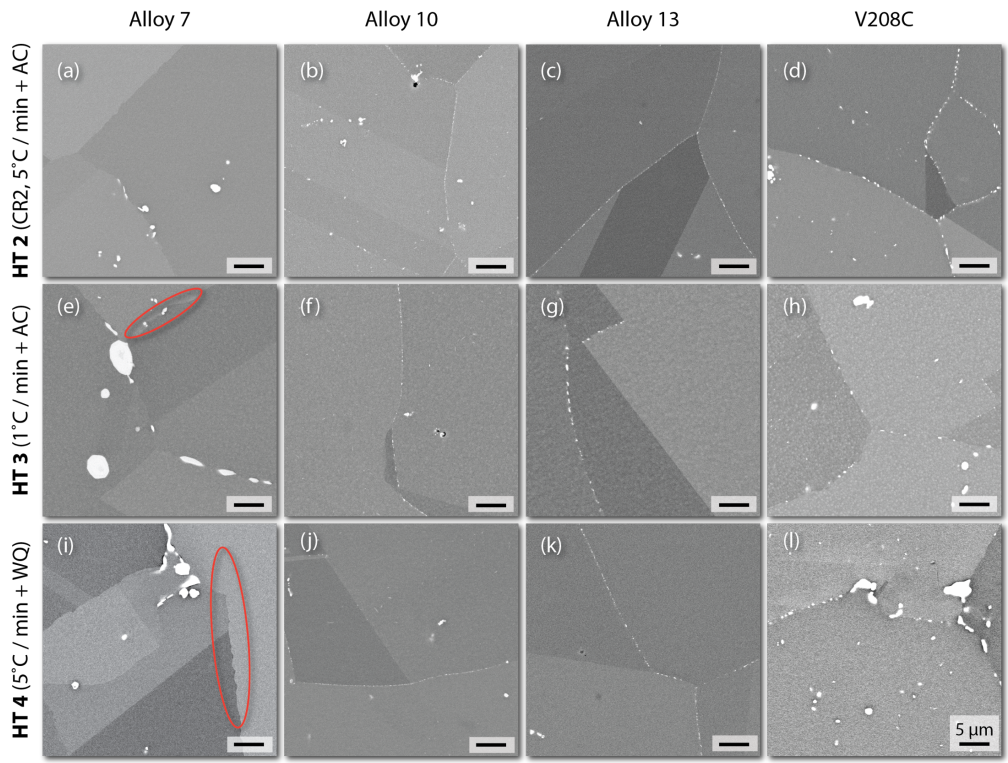

**Figure 11**: Grain boundary observations for four alloys and three heat treatments (varied cooling rate, see Fig. 3). Serrations are highlighted in orange circles. They are only observed when boundary coverage of precipitates is not dense. Backscatter imaging, 20 kV, 10 mm working distance.

We have compared our experimental observations to thermodynamic simulations and observed a stark difference. ThermoCalc predicts precipitation of $M_{23}C_6$, $M_2B$, TCP mu and eta ($Ni_3Ti$). We observe the $M_2B$ boride in this system, but did not once observe $M_{23}C_6$, mu or eta. These phase structures were included in the EBSP template matching. We suggest that there are insufficient accurate observations of these phases in the database we used (TC-Ni8) for ThermoCalc to successfully interpolate to our composition and accurately predict stability. A potentially useful conclusion we draw from the simulations is that $M_2B$ precipitation appears to be inhibited by C, Zr and Ti, while C, B and Zr additions raise the solvus (stabilising) the $M_{23}C_6$ carbide.

This work demonstrates the utility of the PCA-EBSD/EDS method in extracting representative EBSPs and EDS spectra from combined and simultaneous datasets for the purposes of phase classification and observing trends in chemistry. Included within our spatial maps is also the location, size and shape of each phase. In the present manuscript, we have not focussed on these observations, but we note that this is likely important for the relative contributions to high temperature creep performance of these engineering alloys.

Our study used a high voltage and a coarse step size (100 nm) which limits our ability to classify of ultrafine precipitates that are occasionally observed. Adjustment to the sampling geometry, or use of alternative techniques such as TEM or transmission Kikuchi diffraction (TKD), may be required. A combinatorial PCA approach, using VARIMAX, could be used for a TEM diffraction and EDS experiment, but care should especially be taken when considering a VARIMAX rotation for TEM spot diffraction classification as the experimental intra-pattern variance may not necessarily be equal across different structures. Furthermore, even at the relatively loose variance tolerance limit we have used in this work, if a unique signal is only coming from one or two scan points the PCA approach may not see sufficient inter-pattern variance to extract a unique component. We have previously discussed this effect and other PCA artefacts [10].

## 6 Conclusions

From the present study we are able to draw the following conclusions:

1. We have characterised thirteen alloys across C, B, Zr and (low) Ti composition space, as well as two alloys with lower Cr:Al ratio, V208C, and RR1000, to investigate intergranular precipitation. The main alloy series exhibits $M_6C$ and (intragranular) MC carbide precipitation, as well as $M_2B$ borides. We additionally see a NiAl phase in the reduced Cr:Al alloys, but only MC and $M_3B_2$ in RR1000.

2. Interrogation of the chemistries of the grain boundary precipitates in the main alloy reveal them all to be enriched in Mo and W. $M_6C$ draws less Mo than $M_2B$, and appears to see a greater solubility for other elements. The NiAl phase is rich in Ni and Al.

3. Thermodynamic simulations inaccurately predict the precipitation we experimentally observe.

$M_{23}C_6$, TCP mu and eta are all predicted but never observed. The modelling predicts the $M_2B$ boride that we do observe, and that it is destabilised by additions of C, Zr and Ti. This agrees with our lack of boride observation in Zr and Ti enriched alloys.

4. Grain boundary serrations were only achievable at slow cooling rates where grain boundaries did not see spatially dense coverage of $M_6C$ or $M_2B$.

## 7 Statement of work

TPM processed and characterised the presented alloys and prepared the manuscript. Heat treatments were developed by IB, TPM, DD and MH. IB, LRR, TPM, MH, and DD discussed the project direction and selected compositions for investigation. TPM and TBB developed the PCA-EBSD/EDS approach. TBB and AF developed the EBSP refined template indexing procedure. DD and TBB supervised the work.

## 8 Acknowledgements

At Imperial College we acknowledge Thomas Kwok and Xin Xu for assistance in hot rolling, as well as Dafni Daskalaki-Mountanou, Chris Bilsland and Thibaut Dessolier for helpful conversations and participating in developing the PCA approach and characterisations of carbides. TPM, IB, TBB and DD would like to acknowledge support from the Rolls-Royce plc - EPSRC Strategic Partnership in Structural Metallic Systems for Gas Turbines (EP/M005607/1). TPM, TBB and DD acknowledge the Centre for Doctoral Training in Advanced Characterisation of Materials (EP/L015277/1) at Imperial College London. TBB would like to acknowledge the Royal Academy of Engineering for funding his research fellowship. DD acknowledges funding from the Royal Society for his industry fellowship with Rolls-Royce plc.

## Data statement

Data will be uploaded to Zenodo upon article acceptance.